# Toward an Off - Shell 11D Supergravity Limit of M - Theory [1]

Hitoshi NISHINO[2] and S. James GATES, Jr.[3]

*Department of Physics*
*University of Maryland at College Park*
*College Park, MD 20742-4111, USA*

## Abstract

We demonstrate that in addition to the usual fourth-rank superfield $(W_{abcd})$ which describes the on-shell theory, a spinor superfield $(J_\alpha)$ can be introduced into the 11D geometrical tensors with engineering dimensions less or equal to one in such a way to satisfy the Bianchi identities in superspace. The components arising from $J_\alpha$ are identified as some of the auxiliary fields required for a full off-shell formulation. Our result indicates that eleven dimensional supergravity does not have to be completely on-shell. The $\kappa$-symmetry of the supermembrane action in the presence of our partial off-shell supergravity background is also confirmed. Our modifications to eleven-dimensional supergravity theory are thus likely relevant for M-theory. We suggest our proposal as a significant systematic off-shell generalization of eleven-dimensional supergravity theory.

---

[1]This work is supported in part by NSF grant # PHY-93-41926 and by DOE grant # DE-FG02-94ER40854.

[2]E-mail: nishino@umdhep.umd.edu. Also at Department of Physics & Astronomy, Howard University, Washington D.C. 20059, USA.

[3]E-mail: gates@umdhep.umd.edu

## 1. Introduction

There has been a revival of interest in eleven-dimensional (11D) supergravity theory [1]. This revival is occurring within the context of strong/weak duality [2] between 10D type-II superstring theories [3], and 11D supermembrane theory [4], and as a important component of a newly proposed fundamental theory called "M-Theory" [5] suggested to provide a unifying paradigm from which perhaps all superstring and heterotic string theory and various known (as well as unknown dualities) can be derived. If such an underlying theory exists in 11D, we expect its background sector to have a much richer structures than the original 11D supergravity theory [1]. This speculation looks natural, when we recall that 10D superstring theory [3] generated chiral fermions with no cosmological constant *unlike* the original 11D supergravity [1].

As for any significant generalization or modification of 11D supergravity [1], there had been tantalizing speculations on the possibility of higher-derivative terms [6] even prior to the re-birth of string theory. Within superstring theories it is known that higher curvature terms, like the $\alpha'^3 \zeta_{(3)}$ correction from $N = 2A$ superstring to 10D, $N = 2$ supergravity, exist. In the superspace approach [7] for example, the search for higher-order terms *via* a method similar to that developed for superstring corrections to 10D, $N = 1$ supergravity [8][9] at first looks impracticably complicated, due to the $32 \times 32$ matrix representation of the Clifford algebra in 11D, as well as the absence of a dilaton field that could simplify computations [8]. In a component formulation in ref. [10], some generalized Chern-Simons terms were tentatively added to the 11D supergravity Lagrangian [1], but unfortunately the supersymmetric invariance of the total action was not confirmed as expected. There have been some works dealing with auxiliary fields for 11D supergravity [11], but they provide no systematic construction of the off-shell formulation. At the present time, almost twenty years after its initial construction [1], no successful modifications of 11D supergravity with systematic (even perturbative) supersymmetric covariance exist to our knowledge.

We mention, however, an intriguing "glimmer of hope" for the off-shell formulation of 11D supergravity. It was observed that the on-shell superspace formulation of the 11D supergravity theory bore a strange resemblance to the on-shell superspace formulation of the 4D, $N = 2$ supergravity theory [12]. It was also noted that the difference between the on-shell and off-shell versions of 4D, $N = 2$ supergravity was the presence or absence of an auxiliary spinor superfield. On the basis of the similarity between the on-shell theories, it was suggested that an off-shell version of 11D supergravity would necessarily require the presence of a similar spinorial superfield. At that time it was proposed that a future investigation would be undertaken in this direction.

In this paper we take a significant first step toward the non-trivial off-shell generalization of 11D supergravity, motivated by the above indication in 4D, $N = 2$ supergravity. We will prove that there exist a solution of the 11D superspace Bianchi identities in terms of two algebraically independent superfields denoted by $J_\alpha$ and $W_{abcd}$. The latter is the the on-shell field which in a certain limit describes the purely physical and propagating degrees of freedom of 11D supergravity, and it is also an analog of the superfield $W_{\alpha\beta\gamma}$ for 4D, $N = 1$ supergravity [13]. From a geometrical point of view, this multiplet can be called the 11D supergravity "Weyl multiplet". The second superfield $J_\alpha$ is a superfield whose presence implies that the 11D supergravity theory described by our superspace construction is *not* an on-shell construction. It may be thought as the multiplet of auxiliary fields [11] for 11D supergravity. In the following we investigate some of the low dimensional auxiliary fields that it contains. We will not, however, be able to give a complete description of this superfield by the end of this present work.

## 2. Partial Auxiliary Field Structure for 11D, $N = 1$ Supergravity

Our guiding principle in superspace is as usual the satisfaction of the Bianchi identities (BIs):

$$\nabla_{[A} T_{BC)}{}^D - T_{[AB|}{}^E T_{E|C)}{}^D - \frac{1}{2} R_{[AB|c}{}^d (\mathcal{M}_d{}^c)_{|C)}{}^D \equiv 0 \ , \qquad (2.1)$$
$$\frac{1}{24} \nabla_{[A_1} F_{A_2 \cdots A_5)} - \frac{1}{12} T_{[A_1 A_2|}{}^B F_{B|A_3 A_4 A_5)} \equiv 0 \ ,$$

which we call $(ABC, D)$ and $(A_1 \cdots A_5)$-type BIs[4] Our purpose is to satisfy these BIs at engineering dimensions of $d \leq 1$, as the usual fundamental step of solving them [8]. An important guiding principle is to follow the method for the non-minimal 4D, $N = 1$ [12] theory, where a spinor superfield $T_\alpha$ was introduced to generalize the system, and contains some of the auxiliary fields in component approaches. As an 11D, $N = 1$ analog of the $T_\alpha$-superfield, we introduce a spinorial superfield $J_\alpha$, whose first derivative takes the form

$$\nabla_\alpha J_\beta = C_{\alpha\beta} S + i(\gamma^a)_{\alpha\beta} v_a + (\gamma^{ab})_{\alpha\beta} t_{ab} + i(\gamma^{[3]})_{\alpha\beta} U_{[3]} + (\gamma^{[4]})_{\alpha\beta} V_{[4]} + i(\gamma^{[5]})_{\alpha\beta} Z_{[5]} \ . \quad (2.2)$$

Here the subscript $_{[n]}$ stands for the totally antisymmetric indices such as $_{a_1 \cdots a_n}$. In addition to $J_\alpha$ we also introduce a fourth-rank tensor superfield $W_{abcd}$ that is independent of $J_\alpha$ and contains the fields of the purely on-shell theory. In order to go back to the usual on-shell theory, we can just identify $W_{abcd}$ with the fourth-rank field strength $F_{abcd}$, and set $J_\alpha$ simply to zero.

---

[4]For our conventions and notations, see the next section.

We are now ready to present our results for constraints of $d \leq 1$, which constitute the foundation of our modified theory:

$$\begin{aligned}
T_{\alpha\beta}{}^c &= +i(\gamma^c)_{\alpha\beta} \ , \quad F_{\alpha\beta cd} = +\tfrac{1}{2}(\gamma_{cd})_{\alpha\beta} \ , \quad F_{\alpha\beta\gamma\delta} = F_{\alpha\beta\gamma d} = 0 \ , \\
T_{\alpha\beta}{}^\gamma &= -8(\gamma^a)_{\alpha\beta}(\gamma_a)^{\gamma\delta}J_\delta \equiv -8(\gamma^a)_{\alpha\beta}(\gamma_a J)^\gamma \ , \\
T_{\alpha b}{}^c &= +8(\gamma^c \gamma_b J)_\alpha \ , \quad F_{abc\delta} = +12i(\gamma_{abc}J)_\delta \ , \\
T_{\alpha b}{}^\gamma &= +\tfrac{i}{144}(\gamma_b{}^{cdef} + 8\delta_b{}^e \gamma^{def})_\alpha{}^\gamma W_{cdef} \\
&\quad + 8i(\gamma_b)_\alpha{}^\gamma S - 8(\gamma^c \gamma_b)_\alpha{}^\gamma v_c + 8i(\gamma^{cd}\gamma_b)_\alpha{}^\gamma t_{cd} \\
&\quad - 8(\gamma^{cde}\gamma_b)_\alpha{}^\gamma U_{cde} + 12i(\gamma_b{}^{cdef})_\alpha{}^\gamma V_{cdef} - 8(\gamma^{cdefg}\gamma_b)_\alpha{}^\gamma Z_{cdefg} \\
&\quad - 18i(\gamma_b)_\alpha{}^\gamma Q - i(\gamma_b \gamma^{cde})_\alpha{}^\gamma Q_{cde} - 12i(\gamma^{cd})_\alpha{}^\gamma Q_{bcd} \\
&\quad - \tfrac{17i}{12}(\gamma_b \gamma^{cdef})_\alpha{}^g Q_{cdef} - \tfrac{20i}{3}(\gamma^{def})_\alpha{}^\gamma Q_{bdef} \ , \\
R_{\alpha\beta cd} &= +\tfrac{1}{72}(\gamma_{cd}{}^{efgh})_{\alpha\beta}(W_{efgh} + 576 V_{efgh}) + \tfrac{1}{3}(\gamma^{ef})_{\alpha\beta}(W_{cdef} + 576 V_{abcd}) \\
&\quad - 32(\gamma^e)_{\alpha\beta}Q_{cde} - \tfrac{8}{3}(\gamma_{cd}{}^{efgh})_{\alpha\beta}Q_{efgh} - 64(\gamma^{ef})_{\alpha\beta}Q_{cdef} \ , \\
F_{abcd} &= +W_{abcd} + 576\, V_{abcd} \ ,
\end{aligned} \quad (2.3)$$

The $Q$'s are *not* new superfields, but are just products of two $J^\alpha$'s defined by

$$\begin{aligned}
Q &\equiv (\overline{J}J) \equiv J^\alpha J_\alpha \ , \quad Q_{abc} \equiv (\overline{J}\gamma_{abc}J) \equiv J^\alpha (\gamma_{abc})_\alpha{}^\beta J_\beta \ , \\
Q_{abcd} &\equiv (\overline{J}\gamma_{abcd}J) \equiv J^\alpha (\gamma_{abcd})_\alpha{}^\beta J_\beta \ .
\end{aligned} \quad (2.4)$$

Some remarks are now in order. We mention that this set of constraints can *not* be reduced to the original unmodified theory by Cremmer *et al.* [1] by any superfield redefinitions including super-Weyl rescaling [14]. This is critical for our system to really describe a new modification that is not trivially related to the conventional on-shell system.

Note that the superfields $S$, $v_a$, $t_{ab}$, $U_{[3]}$, $V_{[4]}$ and $Z_{[5]}$ are algebraically independent superfields, that play the roles of auxiliary fields in component formulations[5]. In this sense, our modified system already gives an off-shell formulation of 11D, $N = 1$ supergravity. Following 4D, $N = 2$ analysis [12], we also introduce an independent superfield $W_{abcd}$, which is an 11D analog of the tensor superfield $W_{ab}$ of the 4D, $N = 2$ theory. We have determined the constants in (2.3) such that the BI's of $d \leq 1$ are satisfied. To be more

---

[5]In fact, the quantity $U_{[3]}$ is the 11D analog of our 10D $A_{[3]}$-tensor [8].

specific, the BIs at $d \leq 1/2$ require the forms of constraints as

$$T_{\alpha\beta}{}^{\gamma} = + 20(\alpha - 1)\delta_{(\alpha}{}^{\gamma}J_{\beta)} + 2(5\alpha - 9)(\gamma^a)_{\alpha\beta}(\gamma_a J)^{\gamma} - 3(\alpha - 1)(\gamma^{ab})_{\alpha\beta}(\gamma_{ab}J)^{\gamma} ,$$

$$T_{ab}{}^{c} = + 8\delta_b{}^c J_a - 8\alpha(\gamma_b{}^c J)_\alpha , \qquad F_{abc\delta} = +12i(\gamma_{abc}J)_\delta , \qquad F_{abcd} = W_{abcd} + xV_{abcd} ,$$

$$\begin{aligned}T_{ab}{}^{\gamma} = &+ \tfrac{i}{144}\left(\gamma_b{}^{cdef} + 8\delta_b{}^c\gamma^{def}\right)_\alpha{}^\gamma W_{cdef} \\
&+ ia_0(\gamma_b)_\alpha{}^\gamma S + a_1\delta_\alpha{}^\gamma v_b + a_2(\gamma_b{}^c)_\alpha{}^\gamma v_c + ia_3(\gamma_b{}^{cd})_\alpha{}^\gamma t_{cd} + ia_4(\gamma^c)_\alpha{}^\gamma t_{bc} \\
&+ a_5(\gamma_b{}^{cde})_\alpha{}^\gamma U_{cde} + a_6(\gamma^{cd})_\alpha{}^\gamma U_{bcd} + ia_7(\gamma_b{}^{cdef})_\alpha{}^\gamma V_{cdef} + ia_8(\gamma^{cde})_\alpha{}^\gamma V_{bcde} \\
&+ a_9(\gamma_b{}^{cdefg})_\alpha{}^\gamma Z_{cdefg} + a_{10}(\gamma^{cdef})_\alpha{}^\gamma Z_{bcdef} \\
&+ i\xi_1(\gamma_b)_\alpha{}^\gamma Q + \xi_2(\gamma_b\gamma^{cde})_\alpha{}^\gamma Q_{cde} + \xi_3(\gamma^{cd})_\alpha{}^\gamma Q_{bcd} \\
&+ i\xi_4(\gamma_b\gamma^{cdef})_\alpha{}^\gamma Q_{cdef} + i\xi_5(\gamma^{cde})_\alpha{}^\gamma Q_{bcde} ,\end{aligned} \tag{2.5}$$

where $\alpha$, $x$, $a_0$, $\cdots$, $a_{10}$, $\xi_1$, $\cdots$, $\xi_5$ are unknown constants. At $d=1$, in the $(\alpha\beta c, d)$-type BI the symmetric part $(cd)$ in $R_{\alpha\beta cd}$ should be excluded, and this fixes some of the $a$'s. The $(\alpha\beta\gamma,\delta)$-type BI has terms linear in $S$, $v_a$, $t_{[2]}$, $U_{[3]}$, $V_{[4]}$, $Z_{[5]}$, whose coefficients are required to vanish independently due to the algebraic independence of these superfields. Also to be used are the identities in (3.6) in order to have only independent terms with the $U_{[3]}$ and $V_{[4]}$ superfields, as well as the similar identity

$$\tfrac{1}{2}(\gamma^{[a|c})_{(\alpha\beta|}(\gamma^{|b]}{}_c)_{|\gamma)}{}^\delta = -(\gamma_c)_{(\alpha\beta|}(\gamma^c\gamma^{ab})_{|\gamma)}{}^\delta + \tfrac{1}{2}(\gamma^{[a|})_{(\alpha\beta|}(\gamma^{|b]})_{|\gamma)}{}^\delta + (\gamma^{ab})_{(\alpha\beta}\delta_{\gamma)}{}^\delta , \tag{2.6}$$

for $t_{[2]}$-terms.

At this point all the $a$'s are determined uniquely together with $\alpha = 1$:

$$\begin{aligned}&a_0 = +8 , \quad a_1 = -8 , \quad a_2 = +8 , \quad a_3 = +8 , \quad a_4 = -16 , \quad a_5 = +8 , \\
&a_6 = -24 , \quad a_7 = +12 , \quad a_8 = 0 , \quad a_9 = +8 , \quad a_{10} = -40 , \quad \alpha = 1 .\end{aligned} \tag{2.7}$$

The $(\alpha\beta cde)$-type BI has also terms linear in these superfields, vanishing consistently with (2.7), and in particular, the constant $x$ is now fixed to be $x = 576$. Note the curious fact that if we rewrite $T_{ab}{}^\gamma$ using $F_{[4]}$ instead of $W_{[4]}$, then we will find that all the $S, \ldots, Z_{[5]}$-terms in $T_{ab}{}^\gamma$ can be re-combined exactly into a term $-8i(\gamma_b)^{\gamma\epsilon}\nabla_\alpha J_\epsilon$.

The $J^2$-terms can be fixed by the help of Tables 1 through 6 in the next section. All the $\xi$'s are uniquely determined as

$$\xi_1 = -18 , \quad \xi_2 = -1 , \quad \xi_3 = -12 , \quad \xi_4 = -\tfrac{17}{12} , \quad \xi_5 = -\tfrac{20}{3} , \tag{2.8}$$

yielding (2.3).

# 3. Useful Relationships and Identities

Since the computations involved in our analysis are highly technical and lengthy, exposing some crucial identities will be of practical importance. First of all we give notational explanations about the products of our $\gamma$-matrices in 11D. Our basic anticommutator is $\{\gamma^a, \gamma^b\} = +2\eta^{ab} = \text{diag.}(+ - \cdots -)$. Accordingly we have $\gamma^a \gamma^b = +\eta^{ab} + \gamma^{ab}$, where $(\gamma^a \gamma^b)_\alpha{}^\beta \equiv (\gamma^a)_\alpha{}^\gamma (\gamma^b)_\gamma{}^\beta$, $(\gamma^{ab})_\alpha{}^\beta \equiv (1/2)(\gamma^a \gamma^b - \gamma^b \gamma^a)_\alpha{}^\beta$. More generally we define

$$(\gamma^{a_1 \cdots a_n})_\alpha{}^\gamma \equiv \frac{1}{n!} \left[ (\gamma^{a_1})_\alpha{}^{\beta_2} (\gamma^{a_2})_{\beta_2}{}^{\beta_3} \cdots (\gamma^{a_n})_{\beta_n}{}^\gamma + (n! - 1 \text{ perms.}) \right] \quad . \tag{3.1}$$

Accordingly, we have $(\gamma^a)_{\alpha\beta} = (\gamma^a)_\alpha{}^\gamma C_{\gamma\beta} = -C_{\alpha\gamma}(\gamma^a)^\gamma{}_\beta$, *etc.*, where the last expression needs an extra sign as usual [13]. Other important identities are the Fierz identities

$$\delta_{[\alpha}{}^\gamma \delta_{\beta]}{}^\delta = + \frac{1}{16} \left[ C_{\alpha\beta} C^{\gamma\delta} - \frac{1}{6}(\gamma_{abc})_{\alpha\beta}(\gamma^{abc})^{\gamma\delta} + \frac{1}{24}(\gamma_{abcd})_{\alpha\beta}(\gamma^{abcd})^{\gamma\delta} \right] \quad , \tag{3.2a}$$

$$\delta_{(\alpha}{}^\gamma \delta_{\beta)}{}^\delta = - \frac{1}{16} \left[ (\gamma^a)_{\alpha\beta}(\gamma_a)^{\gamma\delta} - \frac{1}{2}(\gamma^{ab})_{\alpha\beta}(\gamma_{ab})^{\gamma\delta} + \frac{1}{120}(\gamma^{[5]})_{\alpha\beta}(\gamma_{[5]})^{\gamma\delta} \right] \quad . \tag{3.2b}$$

We summarize the most useful identities in Tables 1 through 6 below, which will be of great importance, once we have understood the way to use them.[6]

|          | $X_{123}$ | $X_{143}$ | $X_{233}$ | $X_{213}$ | $X_{523}$ |
|----------|-----------|-----------|-----------|-----------|-----------|
| $Y_{23}$ | $-72$     | $0$       | $-40$     | $+16$     | $+1008$   |
| $Y_{43}$ | $+4$      | $-28$     | $+24$     | $-4$      | $-112$    |

Table 1: $\gamma^g$-Multiplication for $Q_{[3]}$

|          | $X_{123}$ | $X_{143}$ | $X_{233}$ | $X_{213}$ | $X_{523}$ |
|----------|-----------|-----------|-----------|-----------|-----------|
| $Y_{13}$ | $-32$     | $0$       | $+160$    | $+96$     | $-1152$   |
| $Y_{33}$ | $+8$      | $0$       | $+40$     | $+16$     | $-432$    |
| $Y_{53}$ | $-4$      | $-28$     | $+16$     | $-4$      | $+32$     |

Table 2: $\gamma^{gh}$-Multiplication for $Q_{[3]}$

---

[6]To our knowledge, these results have never been published in literature for easy access.

|  | $X_{123}$ | $X_{143}$ | $X_{233}$ | $X_{213}$ | $X_{523}$ |
|---|---|---|---|---|---|
| $\widetilde{Y}_{23}$ | $+4$ | $0$ | $+4$ | $0$ | $-72$ |
| $\widetilde{Y}_{43}$ | $+8/3$ | $0$ | $-8/3$ | $-8/3$ | $0$ |
| $Y_{63}$ | $-1/3$ | $0$ | $+1$ | $+2/3$ | $-6$ |
| $Y_{83}$ | $+4$ | $+4$ | $-8$ | $-4$ | $+16$ |

Table 3: $\gamma^{ghklm}$-Multiplication for $Q_{[3]}$

|  | $X_{134}$ | $X_{154}$ | $X_{224}$ | $X_{244}$ | $X_{624}$ |
|---|---|---|---|---|---|
| $Y_{34}$ | $-72$ | $0$ | $+16$ | $-24$ | $-960$ |
| $Y_{54}$ | $+4$ | $-28$ | $-4$ | $+20$ | $+72$ |

Table 4: $\gamma^{g}$-Multiplication for $Q_{[4]}$

|  | $X_{134}$ | $X_{154}$ | $X_{224}$ | $X_{244}$ | $X_{624}$ |
|---|---|---|---|---|---|
| $Y_{24}$ | $-48$ | $0$ | $+64$ | $+144$ | $-1920$ |
| $Y_{44}$ | $+8$ | $0$ | $+16$ | $+56$ | $-320$ |
| $Y_{64}$ | $-4$ | $-28$ | $-4$ | $+12$ | $+8$ |

Table 5: $\gamma^{gh}$-Multiplication for $Q_{[4]}$

|  | $X_{134}$ | $X_{154}$ | $X_{224}$ | $X_{244}$ | $X_{624}$ |
|---|---|---|---|---|---|
| $Y_{14}$ | $-8$ | $0$ | $0$ | $-8$ | $-64$ |
| $\widetilde{Y}_{34}$ | $+12$ | $0$ | $-8$ | $-12$ | $+288$ |
| $\widetilde{Y}_{54}$ | $+4$ | $0$ | $-16/3$ | $-20$ | $+160$ |
| $Y_{74}$ | $-1/3$ | $0$ | $+2/3$ | $-1/3$ | $-56/3$ |
| $Y_{94}$ | $+4$ | $+4$ | $-4$ | $-12$ | $+136$ |

Table 6: $\gamma^{ghklm}$-Multiplication for $Q_{[4]}$

The method to use these tables can be clarified as follows. First, the $X$'s are defined by

$$X_{123} \equiv (\gamma^a)_{(\alpha\beta|}(\gamma^{bc})_{|\gamma)}{}^\delta Q_{abc} , \quad X_{143} \equiv (\gamma^a)_{(\alpha\beta|}(\gamma_a\gamma^{bcd})_{|\gamma)}{}^\delta Q_{bcd} ,$$

$$X_{233} \equiv (\gamma^{ab})_{(\alpha\beta|}(\gamma_a{}^{cd})_{|\gamma)}{}^\delta Q_{bcd} , \quad X_{213} \equiv (\gamma^{ab})_{(\alpha\beta|}(\gamma^c)_{|\gamma)}{}^\delta Q_{abc} ,$$

$$X_{523} \equiv (\gamma^{abcde})_{(\alpha\beta|}(\gamma_{ab})_{|\gamma)}{}^\delta Q_{cde} ,$$

$$X_{134} \equiv (\gamma^a)_{(\alpha\beta|}(\gamma^{bcd})_{|\gamma)}{}^\delta Q_{abcd} , \quad X_{154} \equiv (\gamma^a)_{(\alpha\beta|}(\gamma_a\gamma^{bcde})_{|\gamma)}{}^\delta Q_{bcde} ,$$

$$X_{224} \equiv (\gamma^{ab})_{(\alpha\beta|}(\gamma^{cd})_{|\gamma)}{}^\delta Q_{abcd} , \quad X_{244} \equiv (\gamma^{ab})_{(\alpha\beta|}(\gamma_a{}^{cde})_{|\gamma)}{}^\delta Q_{bcde} ,$$

$$X_{624} \equiv (\gamma^{abcdef})_{(\alpha\beta|}(\gamma_{ab})_{|\gamma)}{}^\delta Q_{cdef} .$$

(3.3)

As is easily seen, the meaning of the indices on $X's$ denote the number of indices on the $\gamma$-matrices and $Q$'s. Note that we need two $\gamma$-matrices for the second factors in $X_{143}$ and $X_{154}$. This is necessary only when $j>k$ for $X_{ijk}$, and $Q_{[k]}$ $(k<j)$ has no free indices. Similarly $Y$'s are defined by

$$Y_{13} \equiv (\gamma^a)_\gamma{}^\delta Q_{gha} , \quad Y_{23} \equiv (\gamma^{bc})_\gamma{}^\delta Q_{gbc} , \quad Y_{33} \equiv (\gamma^{[g|ab})_\gamma{}^\delta Q^{|h]}{}_{ab} ,$$

$$Y_{43} \equiv (\gamma^g\gamma^{abc})_\gamma{}^\delta Q_{abc} , \quad Y_{53} \equiv (\gamma^{gh}\gamma^{abc})_\gamma{}^\delta Q_{abc} , \quad Y_{63} \equiv (\gamma_{[ghkl|}{}^{ab})_\gamma{}^\delta Q_{|m]ab} ,$$

$$\widetilde{Y}_{23} \equiv (\gamma_{[gh|})_\gamma{}^\delta Q_{|klm]} , \quad \widetilde{Y}_{43} \equiv (\gamma_{[ghk|}{}^c)_\gamma{}^\delta Q_{|lm]c} , \quad Y_{83} \equiv (\gamma_{ghklm}\gamma^{abc})_\gamma{}^\delta Q_{abc} ,$$

$$Y_{14} \equiv (\gamma^{[g|})_\gamma{}^\delta Q^{|hklm]} , \quad Y_{24} \equiv (\gamma^{cd})_\gamma{}^\delta Q_{ghcd} , \quad Y_{34} \equiv (\gamma^{def})_\gamma{}^\delta Q_{gdef} ,$$

(3.4)

$$Y_{44} \equiv (\gamma_{[g|}{}^{bcd})_\gamma{}^\delta Q_{|h]bcd} , \quad Y_{54} \equiv (\gamma^g\gamma^{cdef})_\gamma{}^\delta Q_{cdef} , \quad Y_{64} \equiv (\gamma_{gh}\gamma^{abcd})_\gamma{}^\delta Q_{abcd} ,$$

$$Y_{74} \equiv (\gamma^{[ghkl|}{}_{bcd})_\gamma{}^\delta Q^{|m]bcd} , \quad Y_{94} \equiv (\gamma^{ghklm}\gamma^{abcd})_\gamma{}^\delta Q_{abcd} ,$$

$$\widetilde{Y}_{34} \equiv (\gamma_{[gh|}{}^a)_\gamma{}^\delta Q_{|klm]a} , \quad \widetilde{Y}_{54} \equiv (\gamma_{[ghk|}{}^{cd})_\gamma{}^\delta Q_{|lm]cd} ,$$

The *tildes* on $Y$'s in Table 3 or 6 are needed to distinguish them from non-tilded ones in other tables. Note also the special $\gamma$-matrix structures for $Y_{43}$, $Y_{53}$, $Y_{83}$, $Y_{54}$, $Y_{64}$ and $Y_{94}$.

The meaning of Table 1 is now as follows. The first column means that if we multiply $(\gamma^g)^{\alpha\beta}$ by $X_{123}$ with the $\alpha$ and $\beta$-indices contracted, the result will be the linear combination of $Y_{23}$ and $Y_{43}$ with the free indices $g$, $\gamma$ and $\delta$. In particular, Table 1 tells that these terms have the coefficients $-72$ and $+4$, respectively. More explicitly

$$(\gamma^g)^{\alpha\beta}(\gamma^a)_{(\alpha\beta|}(\gamma^{bc})_{|\gamma)}{}^\delta Q_{abc} = -72(\gamma^{bc})_\gamma{}^\delta Q^g{}_{bc} + 4(\gamma^g\gamma^{abc})_\gamma{}^\delta Q_{abc} . \quad (3.5)$$

The advantage of these tables is their compactness to convey so much information for long formulae as above. This can be realized, once we have fixed the convention for the indices on $X$'s and $Y$'s with their contractions. Table 1 through Table 3 are for $Q_{[3]}$, while Tables 4 through Table 6 are for $Q_{[4]}$. The factor $Q_{[n]}$ can be replaced by any arbitrary totally antisymmetric tensor.

Notice the important fact that *not* all of these $X$'s are really independent. As a matter of fact, there exist two identities that relate these $X$'s:

$$X_{523} \equiv +10X_{143} - 24X_{123} - 30X_{213} + 6X_{233} \ , \tag{3.6a}$$

$$X_{624} \equiv +8X_{134} + 2X_{154} - 24X_{224} \ . \tag{3.6b}$$

This can be directly confirmed by the use of the identity such as

$$(\gamma_{ab})_{(\alpha\beta}(\gamma^b)_{\gamma\delta)} \equiv 0 \ . \tag{3.7}$$

We can see that the identities (3.6) are the only relationships among these $X$'s, by solving simultaneous equations out of Tables 1 through 3 or Tables 4 trough 6, respectively for $X_{ij3}$ and $X_{ij4}$.

Another practical usage of these tables can be found even in the $J$-independent sector of the BIs. For example, in the $(\alpha\beta\gamma,\delta)$-type BI we encounter the crucial equation

$$\begin{aligned} &+ \frac{1}{3}(\gamma^{[f|})_{(\alpha\beta|}(\gamma^{|ghk]})_{|\gamma)}{}^\delta + 2(\gamma^a)_{(\alpha\beta|}(\gamma_a\gamma^{fghk})_{|\gamma)}{}^\delta \\ &- (\gamma^{[fg|})_{(\alpha\beta|}(\gamma^{|hk]})_{|\gamma)}{}^\delta - (\gamma^{abfghk})_{(\alpha\beta|}(\gamma_{ab})_{|\gamma)}{}^\delta \equiv 0 \ , \end{aligned} \tag{3.8}$$

but this turns out to be equivalent to (3.6b).

Finally we give here a useful applications of our symbols of $X$'s to important identities. They arise in the $(\alpha\beta\gamma,\delta)$-type BI for products of two $J$'s:

$$\begin{aligned} (\gamma^a)_{(\alpha\beta}J_{\gamma)}(\gamma_a J)^\delta &= \frac{1}{768}\left(+24X_{110} + 4X_{143} - 24X_{123} + X_{154} - 8X_{134}\right) \ , \\ (\gamma^{cd})_{(\alpha\beta}J_{\gamma)}(\gamma_{cd}J)^\delta &= \frac{1}{384}\left(-120X_{110} + 20X_{143} + 24X_{233} - 5X_{154} - 8X_{244}\right) \ , \\ (\gamma^a)_{(\alpha\beta|}(\gamma_{ab}J)_{|\gamma)}(\gamma^b J)^\delta &= \frac{1}{384}\left(+120X_{110} + 8X_{143} + 12X_{123} + X_{154} + 4X_{134}\right) \ , \\ (\gamma^a)_{(\alpha\beta|}(\gamma^a\gamma^{bc}J)_{|\gamma)}(\gamma_{bc}J)^\delta &= \frac{1}{384}\left(-1320X_{110} + 28X_{143} + X_{154}\right) \ , \\ (\gamma^a)_{(\alpha\beta|}(\gamma^b J)_{|\gamma)}(\gamma_{ab}J)^\delta &= \frac{1}{384}\left(-120X_{110} + 8X_{143} - 60X_{123} - X_{154} + 12X_{134}\right) \ , \\ (\gamma^{ab})_{(\alpha\beta|}(\gamma_a J)_{|\gamma)}(\gamma_b J)^\delta &= \frac{1}{384}\Big(-120X_{110} - 20X_{143} + 24X_{213} - 12X_{233} \\ &\quad - 5X_{154} - 4X_{244} + 12X_{224}\Big) \ , \\ (\gamma^{ab})_{(\alpha\beta|}(\gamma_b{}^c J)_{|\gamma)}(\gamma_{ca}J)^\delta &= \frac{1}{384}\Big(+1080X_{110} - 60X_{123} + 24X_{233} + 12X_{134} \\ &\quad - 8X_{244} - 7X_{154}\Big) \ . \end{aligned} \tag{3.9}$$

## 4. Relationship with M-Theory

We now mention a possible important link with M-theory [5], *via* the $\kappa$-symmetry [15] of supermembrane theory [4]. As suggested in the recent literature [16], M-theory seems to be related to the existence of the supermembrane action in a supergravity background. Thus, there is an *a priori* possibility that our putative off-shell supergravity is in conflict with this requirement. This is exactly analogous to the non-trivial consistency confirmation between $\kappa$-symmetry of the 10D Green-Schwarz action and the low-energy *dual* formulation of the heterotic string first performed ten years ago [8]. So a good test of our off-shell supergravity proposal is to see if it is a consistent background for the supermembrane action and likely M-theory.

We recall that the supermembrane action [4] takes the form,[7]

$$I = \int d^3x \left[ +\tfrac{1}{2} \sqrt{-g} g^{ij} \eta_{ab} \Pi_i{}^a \Pi_j{}^b - \tfrac{1}{2} \sqrt{-g} - \tfrac{1}{3} \epsilon^{ijk} \Pi_i{}^C \Pi_j{}^B \Pi_k{}^A A_{ABC} \right] \equiv \int d^3x\, L \ . \quad (4.1)$$

In our off-shell backgrounds (2.3), the $\kappa$-transformation takes the form [15]

$$\delta_\kappa E^\alpha \equiv (\delta_\kappa Z^M) E_M{}^\alpha = (I + \Gamma)^\alpha{}_\beta \kappa^\beta \ , \quad \Gamma \equiv + \frac{i}{6\sqrt{-g}} \epsilon^{ijk} \Pi_i{}^a \Pi_j{}^b \Pi_k{}^c \gamma_{abc} \ , \quad (4.2a)$$

$$\delta_\kappa E^a \equiv (\delta_\kappa Z^M) E_M{}^a = 0 \ , \quad \Pi_i{}^A \equiv (\partial_i Z^M) E_M{}^A \ , \quad (4.2b)$$

for the 11D superspace vielbein $E_M{}^A$ with the coordinates $Z^M$. In the invariance check we adopt the 1.5-order formulation, namely we can always use the algebraic $g_{ij}$-field equation as the embedding condition [4]

$$g_{ij} = \eta_{ab} \Pi_i{}^a \Pi_j{}^b \ . \quad (4.3)$$

Now the $\kappa$-transformation of (4.1) yields new terms containing $T_{c\beta}{}^a$ and $F_{abc\delta}$:

$$\delta_\kappa L = + \sqrt{-g} g^{ij} (\delta_\kappa E^\beta) \left[ i \Pi_i{}^\gamma (\gamma^a)_{\gamma\beta} \Pi_{ja} + \Pi_i{}^c T_{c\beta}{}^a \Pi_{ja} \right]$$
$$- \tfrac{1}{3} \epsilon^{ijk} (\delta_\kappa E^\delta) \Pi_j{}^b \Pi_k{}^a \left[ \tfrac{1}{2} \Pi_i{}^\gamma (\gamma_{ab})_{\gamma\delta} + \Pi_i{}^c F_{abc\delta} \right] \ . \quad (4.4)$$

Using (4.3) and (4.2a), as well as relations under (4.3) such as

$$\tfrac{1}{\sqrt{-g}} \epsilon_i{}^{jk} \Pi_j{}^b \Pi_k{}^c \gamma_{bc} \Gamma = -2i \Pi_i{}^a \gamma_a \ , \quad g^{ij} \Pi_i{}^\gamma \Pi_{ja} \gamma^a \Gamma = + \frac{i}{2\sqrt{-g}} \epsilon^{ijk} \Pi_i{}^\gamma \Pi_j{}^a \Pi_k{}^b \gamma_{ab} \ , \quad (4.5)$$

we see that all the $J$-independent terms cancel each other as the usual supermembrane case [4], while all the $J$-dependent terms are finally re-arranged into two kinds of terms proportional to $J^\alpha \Gamma_\alpha{}^\beta \kappa_\beta$ and $J^\alpha \kappa_\alpha$, which turn out to cancel themselves.

---

[7]The notation here is self-explanatory.

The above discussion constitutes a proof of the fact that our off-shell supergravity formulation (2.4) can also provide a consistent background for the supermembrane action. This is necessary for there to be a link [16] of our off-shell system with M-theory [5].

## 5. Concluding Remarks

In this paper we have given partial but explicit results for the beginning of a complete off-shell formulation of a 11D, $N = 1$ supergravity theory that is consistent with the $\kappa$-symmetry of the supermembrane action, interpreted as the desirable zero-mass limit of M-theory [5]. At this point in our understanding, all the off-shell effects are controlled by the spinorial superfield $J_\alpha$, which is an 11D analog of the superspace formulation of 4D, $N = 2$ supergravity developed in 1980 [12].

Our result is certainly "a counter example" to the conventional myth that there can exist *no* off-shell formulation for 11D supergravity [1] with auxiliary fields. It has long been thought that due to the strict supersymmetry in the maximal dimensions (where the complete on-shell theory contains only gauge fields with no matter fields), there would be no generalization to an off-shell formulation. It is remarkable that to the orders we have checked, we have succeeded in overcoming this taboo by introducing two algebraically independent superfields $W_{abcd}$ and $J_\alpha$. The recent development in M-theory [5] has provided a strong motivation to put an end to such a myth, introducing some auxiliary fields that will be important to accommodate possible higher-order curvature corrections presumably generated from M-theory.

Our results stress the importance of the superspace formulation [13] of supergravity, which still remains as the most powerful tool for supergravity and related subjects even more than twenty years after the first discovery of superspace formulations [17] and supergravity itself [18]. This is because it is always the case both in superstring and supermembrane theories that the critical zero mass sector of a total theory is controlled by supergravity, whose understanding plays a key role in the formulations as field theories with which we can not dispense. Even though superspace formulations remain a tool, it contains still many mysteries that no one has completely mastered.

An important application of our result is its utilization *via* strong/weak duality [2], namely a kind of dimensional reduction into 10D, $N = 2$ supergravity/superstring theory [3] should exist. In particular, we expect the structure of superstring corrections in 10D $N = 2$ system differ markedly from our previous proposals [8][9] from those for the 10D, $N = 1$ superstring. With the reduction of our 11D results to 10D $N = 2$ theory, we at last have a geometrical superspace structure to control the appearance of the $\alpha'^3 \zeta_{(3)}$ terms [19] as the lowest order corrections instead of curvature square terms at $\mathcal{O}(\alpha')$ known to occur in the $N = 1$ theory.

We also mention another attempt by one of the authors [20] in order to accommodate generalized Chern-Simons terms into 11D, $N = 1$ supergravity. In this formulation an eleventh-rank antisymmetric field expected in the M-theory arises naturally. Even though we do not know the direct link of our present off-shell formulation to this Chern-Simons modification [20], we have an increased chance to find that all of these theories are related to each other in the context of M-theory [5].

In closing, we do so on a note of caution also. For although we believe our observation is important, we know of at least two arguments that suggest that there must exist at least one other tensor superfield that will be required to have a completely off-shell formalism. This is to be expected even from the structure of the non-minimal 4D, N = 1 supergravity. There it is known that there are *three* algebraically independent tensors $W_{\alpha\beta\gamma}$, $G_{\alpha\dot\beta}$ and $T_\alpha$. In this work we have introduced the eleven dimensional analog of $W_{\alpha\beta\gamma}$ and $T_\alpha$, so apparently it remains to find the eleven dimensional analog of the $G_{\alpha\dot\beta}$. In future works, these aspects of the eleven dimensional theory will require further study.

We expect our explicit results to open completely new directions to be explored for superstring, super p-brane and supergravity theories. We have increased optimism for resolving long-standing problems in superstring theories such as the dualities and vacuum structures of superstring theory.

The authors are grateful to J.H. Schwarz for stimulating discussions that motivated us to look again at our suggestion made in 1980 [12].

# References

[1] E. Cremmer, B. Julia and J. Scherk, Phys. Lett. **76B** (1978) 409; E. Cremmer and Julia, Phys. Lett. **80B** (1978) 48; Nucl. Phys. **B159** (1979) 141.

[2] E. Witten, Nucl. Phys. **B443** (1995) 85.

[3] *See e.g.*, M. Green, J.H. Schwarz and E. Witten, *"Superstring Theory"*, Vols. I and II, Cambridge University Press (1987).

[4] E. Bergshoeff, E. Sezgin, and P. Townsend, Phys. Lett. **189B** (1987) 75; Ann. of Phys. **185** (1988) 330; E. Bergshoeff and M. de Roo, Phys. Lett. **133B** (1984) 67; E. Bergshoeff, M.J. Duff, C.N. Pope and E. Sezgin, Phys. Lett. **224B** (1989) 71; P.K. Townsend, Phys. Lett. **350B** (1995) 184.

[5] J.H. Schwarz, *"The Power of M-Theory"*, Rutgers preprint, RU-95-68 (Oct. 1995), hep-th/9510086; *"Superstring Dualities"*, Caltech preprint, CALT-68-2019 (Sep. 1995), hep-th/9509148.

[6] L. Alvarez-Gaumé, private communication (1982).

[7] E. Cremmer and S. Ferrara, Phys. Lett. **91B** (1980) 61; L. Brink and P. Howe, Phys. Lett. **91B** (1980) 384.

[8] S.J. Gates, Jr. and H. Nishino, Phys. Lett. **173B** (1986) 46 and 52; Nucl. Phys. **B291** (1987) 205.

[9] H. Nishino, Phys. Lett. **258B** (1991) 104.

[10] P. van Nieuwenhuizen, Stony Brook preprint, Print-84-0737 (Aug. 1984).

[11] A. Van Proeyen, Presented at 2nd Europhysics Study Conf. on Unification of Fundamental Interactions, Erice, Sicily (Oct. 1981), in Erice EPS *"Unification 1981"* 367; A. Van Proeyen, Nucl. Phys. **B196** (1982) 489.

[12] S.J. Gates, Jr., Phys. Lett. **95B** (1980) 305.

[13] S.J. Gates, Jr., M.T. Grisaru, M. Roček and W. Siegel, *Superspace*, Benjamin (1982).

[14] S.J. Gates, Jr. and S. Vashakidze, Nucl. Phys. **B291** (1987) 172.

[15] J. Hughes, J. Liu and J. Polchinski, Phys. Lett. **180B** (1986) 370.

[16] C. Schmidhuber, *"D-brane Actions"* Princeton Univ. preprint PUPT-1585 (Jan. 1996), hep-th 960103.

[17] A. Salam and J. Strathdee, Nucl. Phys. **B76** (1974) 477.

[18] D.Z. Freedman, S. Ferrara and P. van Nieuwenhuizen, Phys. Rev. **D13** (1976) 1324; *ibid.* **D14** (1976) 912.

[19] D. Gross and E. Witten, Nucl. Phys. **B277** (1986) 1; M.T. Grisaru, A.E.M. van de Ven and D. Zanon, Phys. Lett. **173B** (1986) 423; Nucl. Phys. **B277** (1986) 388; Nucl. Phys. **B277** (1986) 409; Phys. Lett. **177B** (1986) 347.

[20] H. Nishino, Maryland preprint, UMDEPP, in preparation (Jan. 1996).